\documentclass[12pt]{article}
=0cm \headsep =0cm

\newcommand{\be}{\begin{equation}}
\newcommand{\ee}{\end{equation}}
\newcommand{\bea}{\begin{eqnarray}}
\newcommand{\eea}{\end{eqnarray}}

\usepackage {latexsym}
\usepackage {graphicx}

\begin{document}

\begin{titlepage}
\title{Lorentz violation effects in asymmetric two brane models: a nonperturbative analysis}

\author{K. Farakos \footnote{kfarakos@central.ntua.gr} \\
          Department of Physics, National Technical University of
       Athens \\ Zografou Campus, 157 80 Athens, Greece}
\date{ }
  \maketitle

\begin{abstract}
We consider the case of bulk photons in a Lorentz violating brane
background, with an asymmetric warping between space and time warp
factors. A perturbative analysis, in a previous work, gave an energy
dependent phase (or group) velocity of light:
$V_{ph}(\omega)=V_{ph}(0)-C_G \:\omega^2 \quad (C_G>0)$, which was
derived up to second order of time independent perturbation theory.
In this paper, we go beyond the perturbative result and we study the
nonperturbative behavior of the phase velocity for larger energies,
by solving numerically an eigenvalue problem for the wave function
of the zero mode (4D photon). In particular we see that
$V_{ph}(\omega)$ is in general a monotonically decreasing function
which tends asymptotically to a final value $V_{ph}(\infty)$. We
compare with the results of perturbation theory and we obtain  a
very good agreement in the range of small energies. We also present
a wave function analysis and we see that in the nonperturbative
sector of the theory (very high energies), the zero mode and the
massive KK modes tend to decouple from matter localized on the TeV
brane.
\end{abstract}
\end{titlepage}

\tableofcontents

\section{Introduction}

Lorentz symmetry is assumed to be an exact symmetry of nature.
However, there are many exotic theories, mainly quantum gravity and
string models, which predict Lorentz violating effects in the high
energy limit, see \cite{Mavromatos:2009xg} and references therein.
Such an effect is an energy dependent velocity of photons, as in the
context of quantum gravity models of space time foam, where the
vacuum behaves like a medium with a nontrivial subluminous
refractive index. Accordingly, different times of arrival are
expected when photons with very high energies, which are emitted
simultaneously from remote astrophysical sources, reach the
detectors of current experiments. Note that there are recent
experimental data of MAGIC and FERMI telescopes, which imply a time
delay of more energetic photons in comparison against lower-energy
ones. However, such a difference may also have a conventional
astrophysical interpretation: for example, photons with different
energies may be  emitted not simultaneously at their sources.

Beyond quantum gravity models, an alternative mechanism which can
produce a nontrivial vacuum refractive index based on brane models
with an asymmetric space-time warping, was proposed.  This mechanism
was studied in \cite{Farakos:2008rv} using time independent
perturbation theory. Here we will study this mechanism in the
nonperturbative regime, by solving numerically the eigenvalue
problem for the wave function of the 4D photon. Such a study is
useful for an analysis of extremely high energy cosmic phenomena in
the context of asymmetric warp models. An example might be ultra
high energy photons with energies higher than $10^{19}$ eV.

Brane world models
\cite{antoniadis,ArkaniHamed:1998rs,Antoniadis:1998ig} are models
with extra dimensions which are used by theorists in order to
address the hierarchy problem. According to this scenario, standard
model particles are assumed to be localized in a three dimensional
brane (our world), while gravitons can propagate in the
multidimensional bulk. Beyond the ADD scenario
\cite{ArkaniHamed:1998rs,Antoniadis:1998ig} where the extra
dimensions are assumed to be large, brane models in which the bulk
space time is warped have been also
proposed~\cite{Randall:1999ee,Randall:1999vf}. In the case of warped
space-time, the extra dimensions could be: (1)  finite, if a second
parallel brane world lies at a finite bulk distance from our
world~\cite{Randall:1999ee} or (2) infinite, if our world is viewed
as an isolated brane, embedded in an (infinite) bulk
space~\cite{Randall:1999vf}. The previously mentioned model with the
two branes, is often called first Randall Sundrum model (RS1-model).
Generalizations of the above generic models, including, for
instance, bulk fields along the extra dimension(s) or higher-order
curvature corrections, have been also considered, see for example
Refs. \cite{Mavromatos:2005yh, Mavromatos:2000az, Giovannini:2001ta,
Farakos:2005hz} and references therein.

We will adopt the following generic ansatz for the metric in five
dimensions \be
ds^2=-\alpha^{2}(z)dt^2+\beta^{2}(z)d\textbf{x}^2+\gamma^2(z) dz^2
\label{asymm} \ee where $z$ parameterizes the extra dimension. In
contrast to the RS-model where the space and time warp factors are
equal, in models with an asymmetric warping we have in general
$\alpha(z)\neq \beta(z)$. Thus, although the induced metric on the
brane (localized at $z=0$ for example) is Lorentz invariant upon
considering  the case $\alpha(0)=\beta(0)$,  the metric of Eq.
(\ref{asymm}) does not preserve 4D Lorentz invariance in the bulk
since $\alpha(z)\neq\beta(z)$ for $z\neq 0$. In such models Lorentz
violation is due to bulk particles which can "feel" the difference
between the space and time warp factors toward the extra dimension.
In the standard brane-world scenario only gravitons are allowed to
propagate in the bulk, hence, in the tree level, Lorentz violation
effects are expected only in the gravitational sector. In Refs.
\cite{Csaki:2000dm, Cline:2001yt} specific asymmetric models predict
a superluminous propagation of gravitons. However, since the
detection of gravitons is still not an experimental fact, we cannot
use this effect in order to set restrictions to asymmetric brane
models.

As we have already mentioned, in Ref. \cite{Farakos:2008rv} an
asymmetric model where photons can freely move between two parallel
branes in a 5D black hole background was considered. A perturbative
analysis, of this model, gave an energy-dependent phase (or group)
velocity of light:
\begin{equation}
V_{ph}(\omega)=V_{ph}(0)-C_G \:\omega^2 \quad (C_G>0) \label{vel}
\end{equation}
and
\begin{equation} V_{gr}(\omega)=V_{gr}(0)-3C_G \:\omega^2 \quad
 \label{velgroup}
\end{equation}
 which was derived up to second order of time independent
perturbation theory. Usually, in the conventional models the only
bulk particles are the gravitons, but in the case of bulk photons
which will be considered here we can set severe constraints to the
free parameters of asymmetric models, see Ref.
\cite{Farakos:2008rv}.

In this paper, we have examined the nonperturbative regime of the
model in Ref. \cite{Farakos:2008rv}, by solving numerically an
eigenvalue problem for the wave function of the zero mode (4D
photon). We found that $V_{ph}(\omega)$ is indeed given by the
perturbative formula of Eq. (\ref{vel}) in the range of small
energies, but there is an inflexion point after which perturbation
theory is not valid, then the phase velocity decreases monotonically
with energy and tends asymptotically to a limiting value
$V_{ph}(\infty)$.

In section 2 we introduce an asymmetric model with bulk photons,
which consists of two branes in a 5D charged black hole background,
and we examine in detail the corresponding junction conditions on
the two branes. In section 3 we present the nonperturbative analysis
for the phase velocity and group velocity of 4D photon, while in
section 4 we study the behavior of the wave function of the zero
mode and the first KK excitation. Finally in section 5, we present
our conclusions and we discus our results in connection with the
MAGIC experiment, and the recent severe restrictions of ultra high
energy cosmic rays on quadratic dispersion relations for the
velocity of light.

\section{Asymmetric two brane models }

\subsection{5D AdS-Reissner-Nordstrom black holes}

We consider an action  which includes 5D gravity, a negative
cosmological constant $\Lambda$, plus a bulk U(1) gauge
field~\cite{Csaki:2000dm}:
\begin{equation}
S=\int d^5 x \sqrt{g}\left(\frac{1}{16 \pi G_5}(R^{(5)}-2\Lambda)-
\frac{1}{4}B^{MN}B_{MN} \right)+\int d^4 x
\sqrt{g^{(+)}_{(br)}}{\cal L}_{m}^{(+)}++\int d^4 x
\sqrt{g^{(-)}_{(br)}}{\cal L}_{m}^{(-)}, \label{matter}
\end{equation} where $G_{5}$ is the five dimensional Newton
constant, and $B_{MN}=\partial_{M}H_N-\partial_{N}H_{M}$ is the
field strength of the U(1) gauge field $H_M$, with $M,N = 0,1 \dots
4$. The four-dimensional terms in the action correspond to matter
fields localized on the two branes of the model, which are located
at $r=r_{+}$ and $r=r_{-}$  $(r_{-}<r_{+})$, and described by two
\emph{perfect fluids}, localized on the two branes, with energy
momentum tensors
\begin{eqnarray}
&& T_{(+)\mu}^{\;\nu}={\rm Diag}(-\rho_{+},p_{+},p_{+},p_{+})\delta(r-r_{+}) \label{matter1}\\
&& T_{(-)\mu}^{\;\nu}={\rm Diag}(-\rho_{-},p_{-},p_{-},p_{-})\delta(r-r_{-}) \label{matter2}
\end{eqnarray}
As we will see in section 2.2, these brane terms are necessary for
the solution of Eqs. (\ref{csaki0}) and (\ref{rns}) below to satisfy
the Israel junction conditions on the two branes.

The corresponding Einstein equations can be written as
\begin{equation}
G_{MN}+\Lambda g_{MN}=8\pi G_5\left(\frac{\sqrt{|g_{(br)}^{(+)}|}}{\sqrt{|g|}} T^{(+)}_{\mu\nu}\delta_{M}^{\mu}\delta_{N}^{\nu}+\frac{\sqrt{|g_{(br)}^{(-)}|}}{\sqrt{|g|}} T^{(-)}_{\mu\nu}
\delta_{M}^{\mu}\delta_{N}^{\nu}+T_{MN}^{(B)} \right)   \label{einstein1}
\end{equation}
where the energy momentum tensor for the U(1) Gauge field is:
\begin{equation}
T_{MN}^{(B)}=B_{MP}B_{N}^{~P}-\frac{1}{4}g_{MN} B_{PS}B^{PS}
\end{equation}
For the metric of the black hole solution we make the ansatz
\begin{equation}
ds^2=- h(r) dt^2+\ell^{-2}r^2d\Sigma^2+h(r)^{-1}dr^2  \label{csaki0}
\end{equation}
where $d\Sigma^2=d\sigma^{2}+\sigma^{2}d\Omega^2$ is the metric of the spatial 3-sections, which
in our case are assumed to have zero curvature. Moreover, $\ell$ is the AdS radius which is equal to
$\sqrt{-\frac{6}{\Lambda}}$.

By solving the Einstein equations (\ref{einstein1}) we obtain:
\begin{equation}
h(r)=\frac{r^2}{\ell^2}-\frac{\mu}{r^2}+\frac{Q^2}{r^4} \label{rns}
\end{equation}
where $\mu$ is the mass (in units of the five dimensional Planck scale) and $Q$ the charge of
the 5D \emph{AdS-Reissner-Nordstrom} black hole.
This, of course, presupposes the existence of extra bulk matter, namely a point-like source with
mass $\mu$ and charge $Q$.
Note that, in the case of nonzero charge $Q$, a non-vanishing component $B_{0r}$ of the bulk
field-strength tensor $B_{MN}$:
\begin{equation}
B_{0r}=\frac{\sqrt{6}}{\sqrt{8\pi G_5}}\frac{Q}{r^3}~,
\end{equation}
is necessary so that the solution satisfies the corresponding
Einstein-Maxwell equations.

\subsection{Junction conditions}

 We can use the 5D AdS-Reissner-Nordstrom black hole solution in order to construct two brane models.
 As a first step, we place two branes, one at the position $r=r_{+}$ (Planck brane) and the other at the
 position $r=r_{-}$ (TeV brane) (note that $r_{+}>r_{-}$). We next assume that for $r<r_{+}$ the 5D
 metric is given by Eq.~(\ref{csaki0}), while for $r>r_{+}$ the metric is given by Eq.~(\ref{csaki0})
 upon the replacement $r\leftrightarrow r_{+}^2/r$. The metric which is obtained in this way is $Z_2$-symmetric
 upon the replacement $r\leftrightarrow r_{+}^2/r$, and the points
 $r_{+}$ and $r_{-}$ correspond to the fixed points of the orbifold structure of the model.

 The next step is to glue the two independent slices of the metric by including two perfect fluid energy
 momentum tensors on both branes, see Eqs. (\ref{matter1}) and (\ref{matter2}) above. Then we have to
 satisfy the junction conditions at the positions $r=r_{+}$ and $r=r_{-}$ (four junction conditions).
 In Refs. \cite{Csaki:2000dm, Cline:2001yt} the junction condition for the corresponding single brane model
 has been derived. If we apply it in our case we take:
 \begin{equation}
 6\sqrt{h(r_{\pm})}=\pm k_{5}^{2} \rho_{\pm} r_{\pm}, \quad 18 h'(r_{\pm})=-k_{5}^{4}(2+3\omega_{\pm})\rho_{\pm} r_{\pm}
 \end{equation}
and after same algebra we obtain
\begin{eqnarray}
\frac{\mu\ell^2}{3 r_{+}^4}&=&(1+\frac{\omega_{+}}{36}k_5^4 \ell^2\rho_{+}^2)=(1+\frac{\omega_{-}}{36}k_5^4\ell^2\rho_{-}^2) \epsilon^4\\
\frac{Q^2\ell^2}{2 r_{+}^6}&=&(1+\frac{1+3\omega_{+}}{72}k_5^4\ell^2\rho_{+}^2)=(1+ \frac{1+3  \omega_{-}}{72}k_5^4\ell^2\rho_{-}^2) \epsilon^6
\end{eqnarray}
where $k_5=8\pi G_5$, and ($\rho_{+}$, $\rho_{-}$) and ($p_{+}$,
$p_{-}$) are the energy densities and pressures on the Planck and
the  TeV brane correspondingly . The equations of state are
parameterized as usual by:
\begin{equation}
\omega_{+}=p_{+}/\rho_{+}, \quad \omega_{-}=p_{-}/\rho_{-}
\end{equation}
Note that $\rho_{+}>0$ (positive tension brane) and $\rho_{-}<0$ (negative tension brane).
The parameter $\epsilon$ is defined as:
\begin{equation}
\epsilon=\frac{r_{-}}{r_{+}}
\end{equation}
In order to address the hierarchy problem, in a similar way with
that of RS1-model, we have to choose a very large ratio
$\epsilon=10^{-16}$. Now, if we define the parameters
\begin{eqnarray}
 \bar{\rho}_{+}= \frac{k^2_{5} \ell}{6}  \; \rho_{+}, \quad \bar{\rho}_{-}=\frac{k^2_{5} \ell}{6}   \; \rho_{-}\\
 \bar{\mu}=\frac{\mu\ell^2}{3 r_{+}^4}\epsilon^{-4}, \quad \bar{Q}^2=\frac{Q^2\ell^2}{2 r_{+}^6}\epsilon^{-6}
\end{eqnarray}
we obtain that
\begin{eqnarray}
\bar{\mu}&=&(1+\omega_{+}\bar{\rho}_{+}^2)\epsilon^{-4}=(1+\omega_{-}\bar{\rho}_{-}^2)\label{first} \\
\bar{Q}^2&=&(1+\frac{1+3\omega_{+}}{2}\bar{\rho}_{+}^2)\epsilon^{-6}=(1+ \frac{1+3  \omega_{-}}{2}\bar{\rho}_{-}^2) \label{second}
\end{eqnarray}
In what follows we will consider that $0<\bar{\mu}<<1$ and
$0<\bar{Q}^2<<1$, as our purpose is to construct two brane models
that are described by asymmetric metrics which are linearized
perturbations around the RS1 metric. See also Eq. (\ref{deviation})
for the perturbation $\delta h$ below. This implies that $r_{+}$,
which is the radius that determines the position of the Planck brane
in the bulk, is (comparatively) a very large quantity \footnote{Also
$r_{-}$ is a very large quantity because
$\bar{\mu}=\frac{\mu\ell^2}{3 r_{-}^4}$ and
$\bar{Q}^2=\frac{Q^2\ell^2}{2 r_{-}^6}$ are assumed to be very small
numbers.}. In particular, we have to satisfy both the inequalities $
r_{+}^2 \epsilon^{2}  \gg \sqrt{\mu} \ell$ and $r_{+}^3 \epsilon^3
\gg Q \ell$.

By solving Eqs. (\ref{first}) and (\ref{second})
(they are four algebraic equations) we find the energy densities:
\begin{eqnarray}
\bar{\rho}_{+}^2&=&1-3 \bar{\mu}\; \epsilon^4+2 \bar{Q}^2 \epsilon^6 \label{en1}\\
\bar{\rho}_{-}^2&=&1-3 \bar{\mu }+2 \bar{Q}^2 \label{en2}
\end{eqnarray}
and the equation of state parameters
\begin{eqnarray}
w_{+}&=&-1-2 \bar{\mu}\; \epsilon^4+2 \bar{Q}^2 \label{w1}  \epsilon^6 \label{den1}\\
w_{-}&=&-1-2 \bar{\mu }+2 \bar{Q}^2 \label{w2} \label{den2}
\end{eqnarray}
We see that the energy densities $\bar{\rho}_{+}$,  $\bar{\rho}_{-}$
and the state factor parameters $w_{+}$, $w_{-}$ depend only on the
constants $\bar{\mu}$ and $\bar{Q}^2$ and the hierarchy  parameter
$\epsilon$. Note that for $\omega_{+}=-1$ and $\omega_{-}=1$ we
obtain the first RS-model.

The equation of state parameters should respect the null energy condition $\omega>-1$,
hence if we demand $\omega_{+}\geq-1$ and $\omega_{-}\geq-1$  we obtain the following constraint :
\begin{equation}
\bar{\mu}\leq \bar{Q}^{2} \epsilon^2 \label{constr0}
\end{equation}
This equation means that we can choose the parameters $\bar{\mu}
\;(>0)$ and $\bar{Q}^2 \; (>0)$ arbitrarily insofar as they satisfy
the constraint of Eq. (\ref{constr0}). Note that between the two
branes there are no horizons as the parameters $\bar{\mu}$ and
$\bar{Q}^2$ are assumed to be very small, and the positions $r_{+}$
and $r_{-}$ of the two branes are very large.

Finally, we would like to stress the fact that, although this
problem has been considered previously in the literature, see Refs.
\cite{Csaki:2000dm, Cline:2001yt}, the results for the energy
densities (Eqs. (\ref{en1}) and (\ref{en2})) and the state factors
(Eqs. (\ref{den1}) and (\ref{den2})) are presented for the first
time in the present paper.

\subsection{5D AdS-Reissner-Nordstrom Solution as a linearized perturbation around the Randall-Sundrum metric}

To write the 5D AdS-Reissner-Nordstrom solution as a
\emph{linearized perturbation} around the RS metric, we perform the
following change of variables $r \to z(r)$ in Eq. (\ref{csaki0}):
\begin{eqnarray}
r&=& r_{+} e^{-k\;z}~, \quad {\rm for}~  z>0 \nonumber \\
r&=& r_{+} e^{k\;z}~, \quad  {\rm for}~ z<0~,
\end{eqnarray}
If we rescale $x_{\mu}\rightarrow \frac{r_{+}}{\ell}x_\mu \quad
(\mu=0,\dots ,3)$,  we obtain: \bea ds^2=-a^2(z) h(z)
dt^2+a^2(z)d\textbf{x}^2+h(z)^{-1}dz^2  \label{csaki} \eea where
$a(z)=e^{-k|z|}$, and $k=\ell^{-1}$ is the inverse $AdS_5$ radius.
For the function $h(z)$ we obtain:
\begin{equation}
h(z)=1-\delta h(z), \quad \delta h(z)=3\: \bar{\mu}\: \epsilon^4 \:e^{4 k|z| }-2\:\bar{Q}^2 \: \epsilon^6\: e^{6k|z|} \label{deviation}
\end{equation}
The positions of the branes which are located at $r_{+}$ and
$r_{-}=r_{+} \epsilon$ in the original coordinate system, are
determined in the new coordinate system by the equations $z=0$ and
$z=z_c$ correspondingly, where $\epsilon=e^{-k z_c}$ ($r_c=z_c/\pi$
is radius of the compact extra dimension). Note that the large
hierarchy $\epsilon\sim10^{-16}$ is achieved if we choose
$z_{c}\simeq 37$.  In addition, we will assume that $|\delta
h(z)|\ll 1$ in the interval $0<z<z_c$, or equivalently we adopt that
$\delta h(z)$ is only a small perturbation around the RS-metric. We
shall use the term Planck brane for the positive tension brane at
the position $z=0$, and the term TeV brane for the negative tension
brane, at $z_c$.

\subsection {Bulk photons in asymmetric two brane models}

In this section we will study the case of a 5D massless $U(1)$ gauge
boson $A_{N}$ in the background of an asymmetrically warped solution
of the form of Eq. (\ref{csaki}). We stress that the gauge field
$A_{N}$ must not be confused with the gauge field $H_{N}$,
introduced in the previous section. As we will see later, we will
identify the zero mode of $A_{N}$ with the standard four dimensional
photon. On the other hand $H_{N}$ is an additional bulk field which
does not interact with the charged particles on the brane. The
equation of motion for $A_{N}$ reads: \be
\frac{1}{\sqrt{g}}\partial_{M}\left(\sqrt{g}
g^{MN}g^{RS}F_{NS}\right)=0~, \label{photon} \ee with
$F_{NS}=\partial_{N}A_S-\partial_{S}A_N$, and $N,S= 0,1,\dots 5$. In
the background metric of Eq. (\ref{csaki}), Eq.~(\ref{photon})
gives: \bea -\partial_z(a^{2}(z) h(z) \partial_z
A_j)-\nabla^{2}A_j+\frac{1}{h(z)}\partial_{0}^{2}A_j=0, \quad
j=1,2,3~, \eea where we have assumed the Coulomb gauge condition:
\be \vec{\nabla }\cdot \vec{A}=0, \quad A_{0}=0,\quad A_{z}=0 ~,
\label{coulomp} \ee which is suitable for the case of a Lorentz
violating background. On setting in Eq. (\ref{photon}): \be
A_j(x,z)=e^{i p \cdot x}\chi_j(z), \quad
p_{\mu}=(-\omega,\textbf{p}) \ee we obtain \be
-\partial_{z}\left\{a^{2}(z) h(z)\partial_z\chi\right\}
+\left\{\textbf{p}^2-\frac{\omega^2}{h(z)}\right\}\chi=0
\label{sch1} \ee where for brevity we have dropped the index $j$
from $\chi$. Note that the \emph{spectrum} of Eq.~(\ref{sch1}) is
\emph{discrete}, due to the orbifold boundary
conditions~\cite{Randall:1999ee}, $\chi'(0)=0$ and $\chi'(z_c)=0$
(where the prime denotes a $z$-derivative).

We would like review here briefly the spectrum in the case of
RS1-model ($\delta h=0$), which consist of a zero mode plus an
infinite tower of massive KK modes. It suffices to mention that the
nonzero eigenvalues are:
\begin{equation}
m^{(0)}_{n}=x_{n}\:k\:e^{-k z_{c}}, \: n=1,2,3,... \label{eigen}
\end{equation}
where $x_{n}$ are the roots of the zeroth order Bessel function $J_{0}(x_n)=0$. On adopting $k z_{c}\sim 12$,
which is
the standard choice in order to connect electroweak (ew) and Planck scales ($M_P=e^{k z_c} m_{ew} $) in a RS
framework~\cite{Randall:1999ee},
one obtains that:
\begin{equation}
m^{(0)}_{n}\sim {\rm TeV}~, \qquad n = 1,2,\dots ~.
\label{mtev}
\end{equation}
The corresponding eigenfunctions are:
\begin{eqnarray}
&&\chi_{0}^{(0)}=\frac{1}{N_0}, \quad \quad \quad \quad \quad \quad N_0=\sqrt{z_c}  \label{eigen1}\\
&&\chi_{n}^{(0)}=\frac{1}{N_n} e^{k z}J_{1}(\frac{m^{(0)}_n}{k}e^{kz}), \quad N_n=\frac{e^{k z_c}}{\sqrt{2 k}}J_1(x_n) \label{eigen2}
\end{eqnarray}
where the coefficients $N_0, \; N_n$ are defined by the normalization condition:
\begin{equation}
\int_{0}^{z_c} \chi_{n}^{(0)}(z)\chi_{m}^{(0)}(z) dz=\delta_{mn}
\end{equation}

\begin{figure}[h]
\begin{center}
\includegraphics[scale=0.5,angle=0]{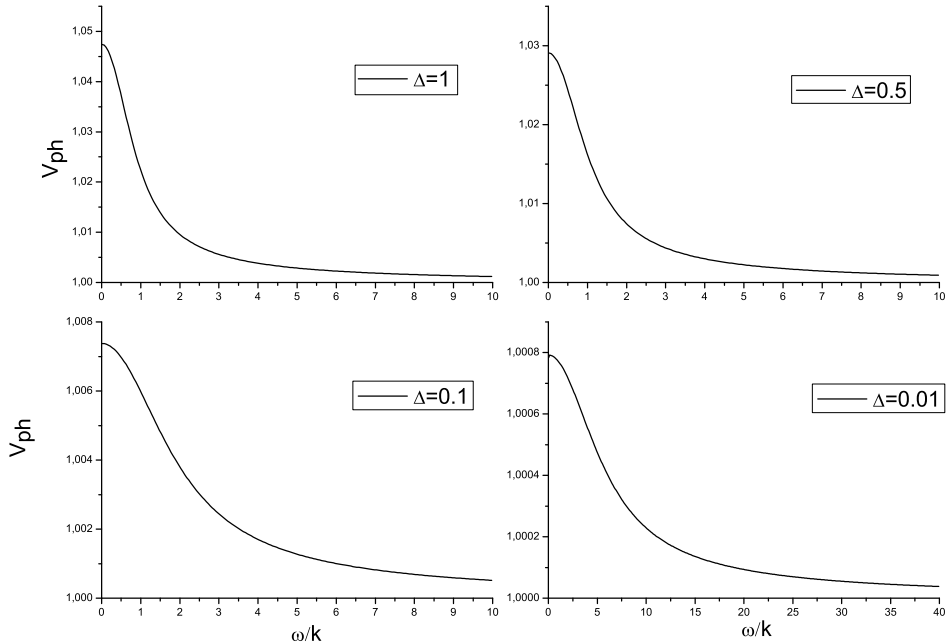}
\end{center}
\caption{\small{The phase velocity of the photon $V_{ph}$ as function of $\omega/k$
for $k z_{c}=2$ and $\Delta=1, \:0.5,\;0.1,\: 0.01$. We see that as $\Delta$
decreases the perturbative region of $V_{ph}$ spreads towards larger energies $\omega/k$}. } \label{1}
\end{figure}
\begin{figure}[h]
\begin{center}
\includegraphics[scale=0.4,angle=0]{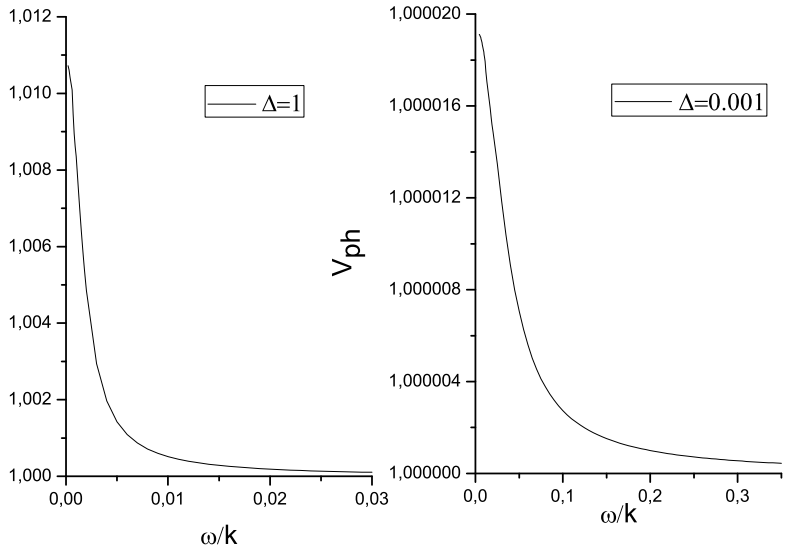}
\end{center}
\caption{\small{The phase velocity of the photon $V_{ph}$ as
function of $\omega/k$ for $k z_{c}=8$ and $\Delta=1, 0.001$. We see
that even for larger hierarchies $kz_c$ ($\epsilon=e^{-kz_c}$) the
qualitative features of the phase velocity do not change.}}
\label{2}
\end{figure}
\begin{figure}[h]
\begin{center}
\includegraphics[scale=0.4,angle=0]{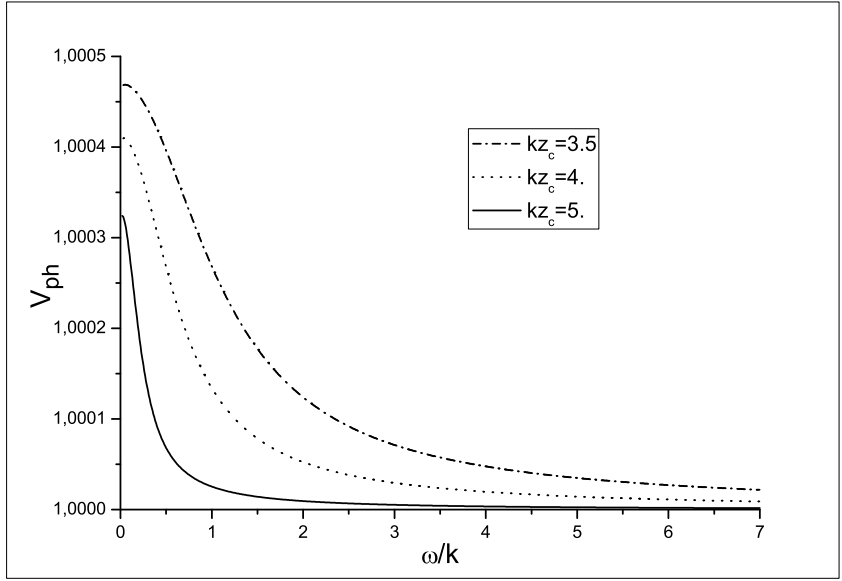}
\end{center}
\caption{\small{The phase velocity of the photon $V_{ph}$ as
function of $\omega/k$  for $k z_{c}=3.5,4,5$ and $\Delta=0.01$. As
$kz_c$ increases the perturbative region shrinks in the low energy
range $\omega/k<<1$, and as a result the velocity $V_{ph}$ tends to
its limiting value in a much faster way.}} \label{3}
\end{figure}
\begin{figure}[h]
\begin{center}
\includegraphics[scale=0.4,angle=0]{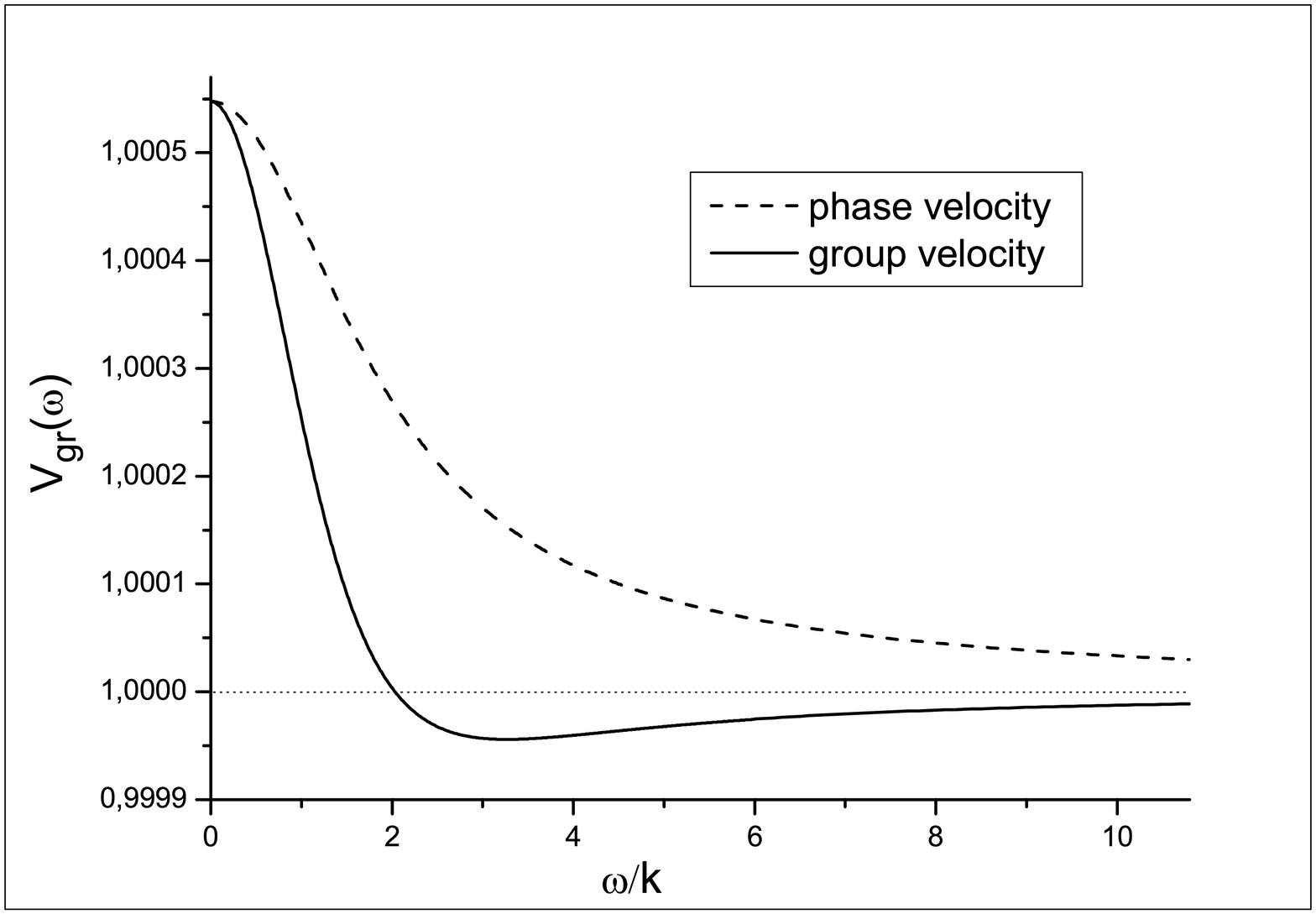}
\end{center}
\caption{\small{The phase and the group velocity of the photon ($V_{ph}$ and $V_{group}$ correspondingly)
 as a function of $\omega/k$ for $k z_{c}=3$ and $\Delta=0.01$.}} \label{4}
\end{figure}

\section{Phase and group velocity of photons: a nonperturbative analysis and a
comparison with perturbation theory}

If we introduce the dimensionless variable $y=kz$ in Eq. (\ref{sch1}) we obtain
\be
\partial^2_y\chi+\left\{-2+\frac{h'}{h}\right\}\partial_y\chi-\frac{1}{a^2 h}\left\{\left(\frac{\textbf{p}}{k}\right)^2-\frac{1}{h}\left(\frac{\omega}{k}\right)^2\right\} \chi=0
 \label{schn}
\ee where \be h(y)=1-\delta \: e^{4y} (3 \: \tilde{c}_{a}-2 \:
e^{2y} ), \quad  \delta=\bar{Q}^2 \epsilon^6, \quad
\tilde{c}_{a}=\frac{\bar{\mu}}{\bar{Q}^{2} \epsilon^2} \ee and the
boundary conditions now read: \be \chi'(0)=0, \quad \chi'(k z_c)=0
\label{bound} \ee Note that $0<\tilde{c}_{a}\leq 1$ if we demand the
null energy condition to be satisfied, see section 2.2 and Eq.
(\ref{constr0}) above. We have also checked that the value of
$\tilde{c}_{a}$ has not a significant impact for our numerical
analysis, hence in what follows we will assume that
$\tilde{c}_{a}=1$, which is the case where $\omega_{+}=-1$, see Eq.
(\ref{den1}) above. Now, it is convenient to introduce a new
parameter \be \Delta=\bar{Q}^2=\delta \: \epsilon^{-6} \ee which can
help us to estimate where perturbation theory fails as an
approximation for solving Eq. (\ref{schn}). In particular, even for
$\Delta<<1$, we can trust perturbation theory only if the energy of
the photon $\omega$ is relatively small. For larger energies
$\omega$ the term \footnote{We have used that $h=1-\delta h$, hence
the term  $\frac{\omega^2}{h}$ in Eq. (\ref{sch1}) can be written as
$\omega^2(1+\delta h)=\omega^2+\omega^2\delta h $. The term
$\omega^2  \delta h $ in the previous equation reveals the
perturbative nature of Eq. \ref{sch1} as $\delta h \approx
\Delta\ll1$. However, for large energies $\omega \approx
1/\sqrt{\delta h}$ perturbation theory breaks down.} $\omega^2\delta
h$ ($\delta h=1-h\simeq \Delta \ll 1$) in Eq. (\ref{sch1}) (or in
Eq. (\ref{schn})) cannot be assumed  small. In this case
perturbation theory breaks down and a nonperturbative approach is
necessary.

The nonperturbative analysis of this section is reduced to an
eigenvalue problem of the second order differential equation
(\ref{schn}) with the boundary conditions (\ref{bound}). We will
consider that $|\textbf{p}|$ is fixed and will try to determine the
energy $\omega$ in order to satisfy the boundary conditions of Eq.
(\ref{bound}). Note that there is an infinite tower of energies
$\omega_{n}$ (n=0,1,2..) which are solutions of the above mentioned
eigenvalue problem for given momentum $|\textbf{p}|$. The first
eigenvalue $\omega_0$ corresponds to the zero mode and the remaining
eigenvalues $\omega_{n}$ ($n\neq 0$) to the massive KK excitations.
Thus, we kept $|\textbf{p}|$ fixed and we integrated numerically
\footnote{For the numerical solving of the eigenvalue problem we
have used mathematica programming.}, for a number of values of
$\omega$ ($\omega>|\textbf{p}|$) separated by a relatively small
constant step $\delta \omega$, Eq. (\ref{schn}) with the initial
condition $\chi(0)=1, \; \chi'(0)=0$. We observed that as we
increased $\omega$, with the constant step $\delta \omega$, the
derivative  $\chi'(kz_c)$ changed sign for first time (from positive
to negative), in an interval of energies $[\omega_a, \omega_a+\delta
\omega ]$. In this interval the first eigenvalue $\omega_0$ can be
determined by bisection method. The infinite tower of energies
$\omega_n$ $(n\neq0)$, can be determined in the same way.

The phase velocity
\begin{equation}
V_{ph}=\omega/|\textbf{p}|
\end{equation}
of the photon (zero mode) as a function of $\omega$ (measured in
units of $k$) has been plotted in Fig.\ref{1}, Fig.\ref{2} and
Fig.\ref{3}, for several values of $\Delta$ and $k z_c$
($\epsilon=e^{-kz_c}$). As we see, the phase velocity is a
monotonically decreasing function which tends asymptotically to a
constant value, seemingly equal to one independently from the
parameters $\Delta$ and $kz_c$.

In Fig. \ref{4} we have plotted the group and phase velocity as a
function of $\omega/k$
\begin{equation}
V_{gr}=\frac{d\omega}{d|\textbf{p}|}
\end{equation}
We can observe that they have a very similar behavior. However, in
contrast with the phase velocity, the group velocity becomes smaller
than unity, then it increases and tends rapidly to unity. Also, as
we see in Fig. \ref{4}, the phase and the group velocity are equal
in the low energy limit as it is expected. It is well known that the
phase and group velocity in vacuum should be identical, however for
larger energies we observe significant differences because the
vacuum behaves as a medium with a non-trivial refractive index. The
perturbative formulas for the phase and group velocity, of Eqs.
(\ref{vel}) and (\ref{velgroup}), agree with our numerical analysis
in the low energy limit, as they give the same value for the group
and phase velocity for zero energy, decreasing quadratically with
energy, while for larger energies outside the perturbative region
they behave differently.
\begin{figure}[h]
\begin{center}
\includegraphics[scale=0.4,angle=0]{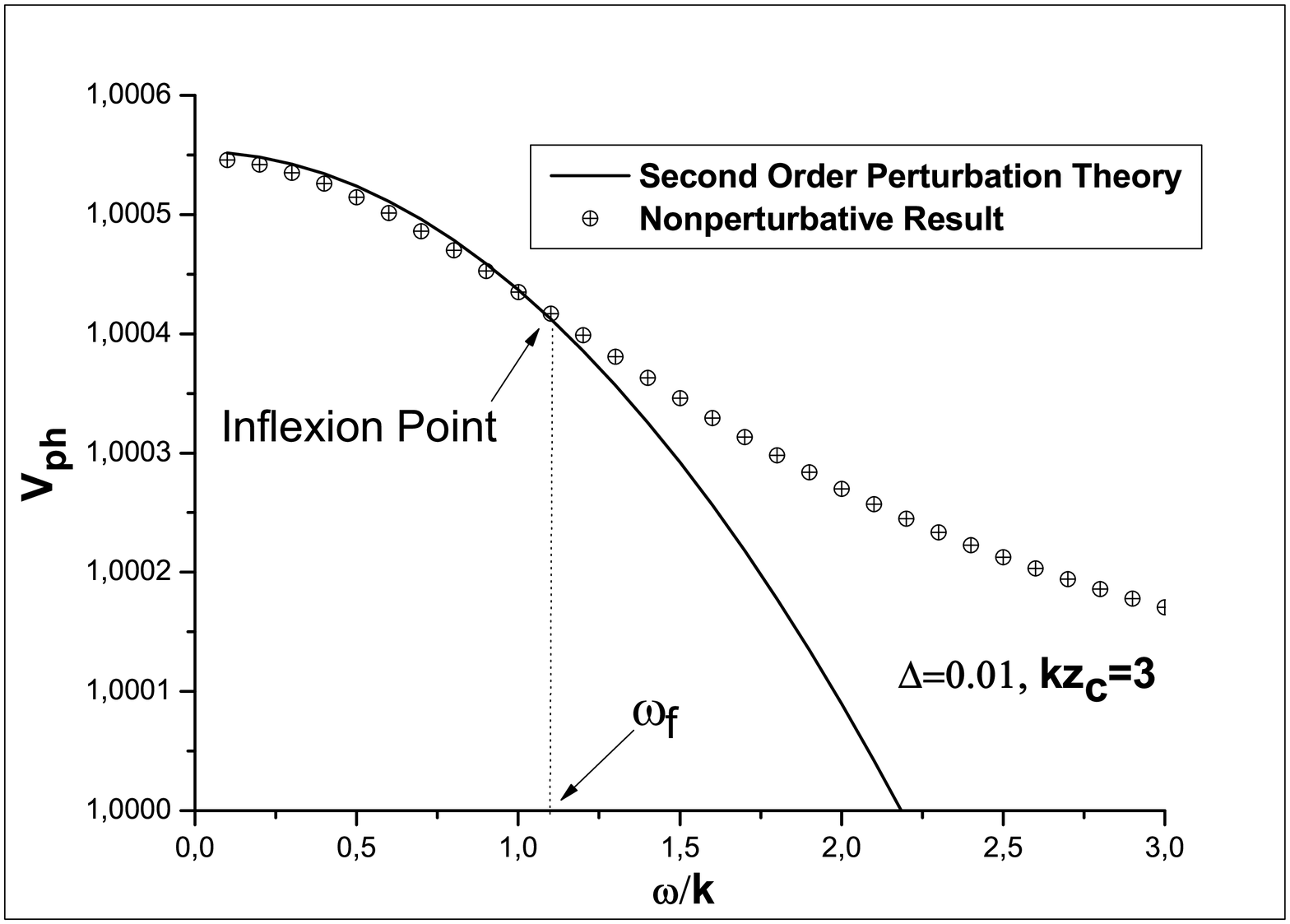}
\end{center}
\caption{\small{The phase velocity of the photon $V_{ph}$ as
function of $\omega/k$  for $k z_{c}=3$ and $\Delta=0.01$. The
discrete points are the nonperturbative results, and the continuous
line corresponds to perturbation theory:
$V_{ph}(\omega)=1.00055-0.00012\: (\omega/k)^2$, where the
coefficients have been computed by the formulas of Eqs. (2.30) and
(2.31) in Ref. \cite{Farakos:2008rv}. We see that the inflexion
point $\omega_f\simeq 1.1\: k$ separates the perturbative from the
nonperturbative sector of the theory}.} \label{5}
\end{figure}

Especially in Fig.\ref{5}, we observe that there is an inflexion
point $\omega_f$ which separates the perturbative from the
nonperturbative sector of the theory. In the perturbative sector
($\omega<\omega_f$) we expect that
\begin{equation}
V_{ph}(\omega)=V_{ph}(0)-C_G \: \omega^2, \quad (C_G>0)  \label{perdesp}
\end{equation}
This formula has been derived in Ref. \cite{Farakos:2008rv} by using second
order time independent perturbation theory. The parameters $V_{ph}(0)$ and $C_G$
are given by the formulas of Eqs. (2.30) and (2.31) in \cite{Farakos:2008rv},
These formulas are suitable for numerical computations, if the parameters $\Delta$ and $kz_c$ are known.

We also see, in Fig.\ref{5}, that when $\omega$ crosses the
inflexion point $\omega_f$ the rate of decreasing of the phase
velocity gets smaller and the phase velocity possesses an asymptotic
value equal to one. Note also, that the perturbative range of the
phase velocity increases (or the inflexion point $\omega_f$ is
displaced towards the right direction in the figures) for smaller
values of the parameter $\Delta$, as we see in Fig.\ref{1} and
Fig.\ref{2}.

At this point, we would like to stress that our numerical analysis
is restricted to an unrealistic range of the parameter space of
$\Delta$ and $kz_c$, which is far beyond the physically interesting
case of $k z_c\simeq 37$. However, it is reasonable that we are not
in position to perform numerical computations in this case, as it
demands great accuracy. This is mainly due to the extremely large
values of the exponential $e^{6kz}$, that appear in Eq.
(\ref{deviation}).

Now, if we compare Fig.\ref{1} for $kz_c=2$ and Fig.\ref{2} for
$kz_c=8$, we see that the qualitative features of the phase velocity
as a function of $\omega$ are unchanged, although there is a
significant difference between the corresponding hierarchies
$\epsilon_1=e^{-2}\sim 10^{-1}$ and $\epsilon_2=e^{-8}\sim 10^{-4}$.
We have also performed computations for larger $kz_c=10$
($\epsilon_3 \sim 10^{-5}$) and we have confirmed the same behavior
for the phase velocity. We also see that this behavior is
independent from the parameter $\Delta$, which determines the
perturbative range of our model. Accordingly, in the conclusions we
will consider an extrapolation assuming that the qualitative
behavior of the phase velocity is also valid for $\epsilon \sim
10^{-16}$ which is the physically interesting case.

Finally in Fig. \ref{3} we see that as $kz_c$ increases the
perturbative range of the phase velocity shrinks near the origin,
where  $\omega/k\ll1$. This behavior is reasonable as $\omega$ in
this figure is measured in units of $k$ (or in units of Planck
scale) and, in the case of realistic values of $k z_c\simeq 37$, the
point where we have the breakdown of perturbation theory is expected
to be several orders of magnitude smaller than the Planck scale (we
will give an estimate of this point in conclusions).

\section{Wavefunction analysis}

\begin{figure}[h]
\begin{center}
\includegraphics[scale=0.4,angle=0]{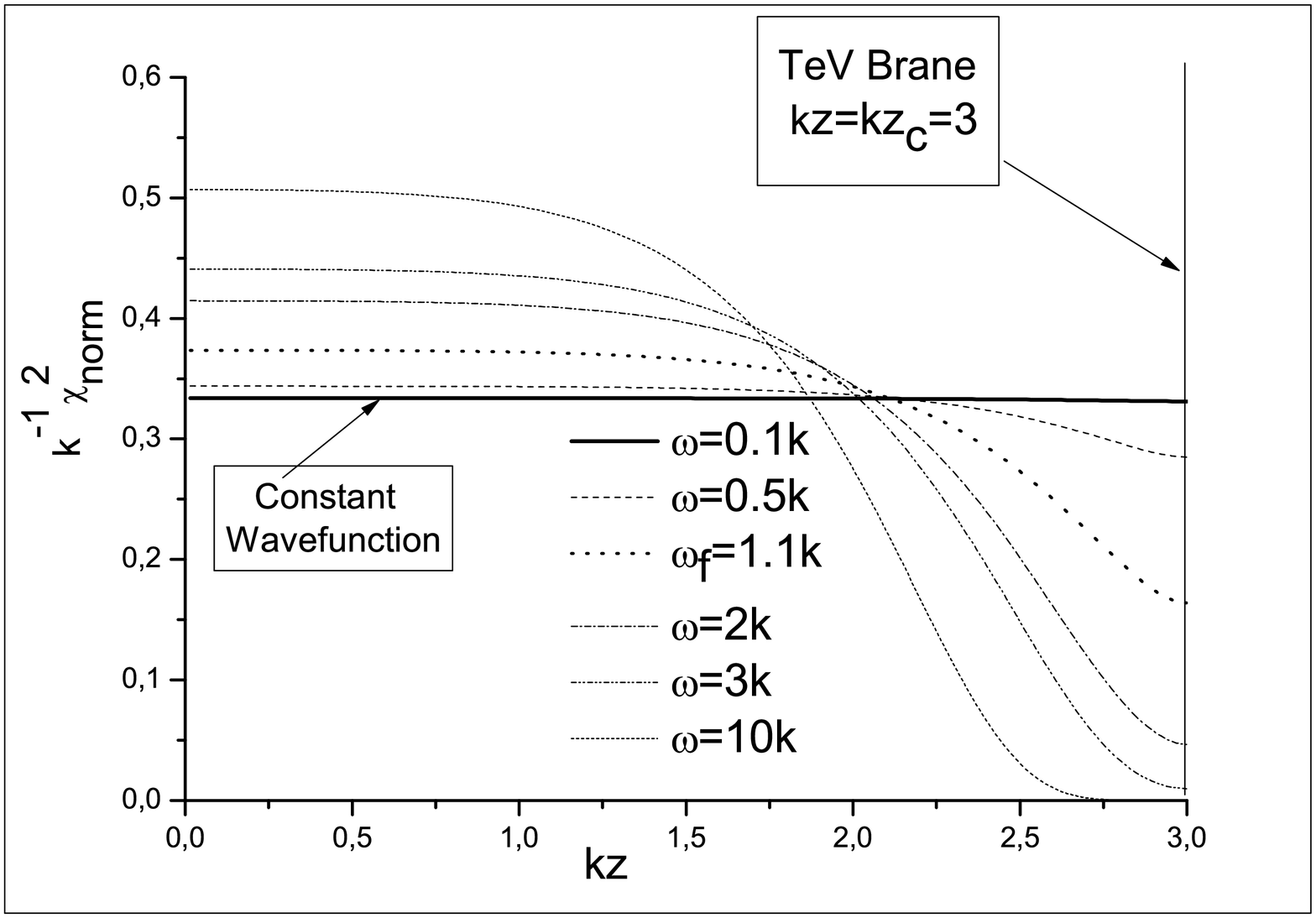}
\end{center}
\caption{\small{The square of the normalized wave function
$k^{-1}\chi_{norm}^2$ of the zero mode (4D photon) as a function of
$kz$ for $k z_{c}=3$, $\Delta=0.01$ and $\omega=0.1k,
\:0.5k,\:1.1k,\:2k,\:3k, \:10k$. We see that the value of the wave
function on the TeV brane (which is proportional to the coupling of
the zero mode with the localized matter on the TeV brane) tends to
zero as the energy of the zero mode increases. On the other hand, on
the Planck brane, we observe that the value of the wave function
increases with the energy.}} \label{6}
\end{figure}

\begin{figure}[h]
\begin{center}
\includegraphics[scale=0.4,angle=0]{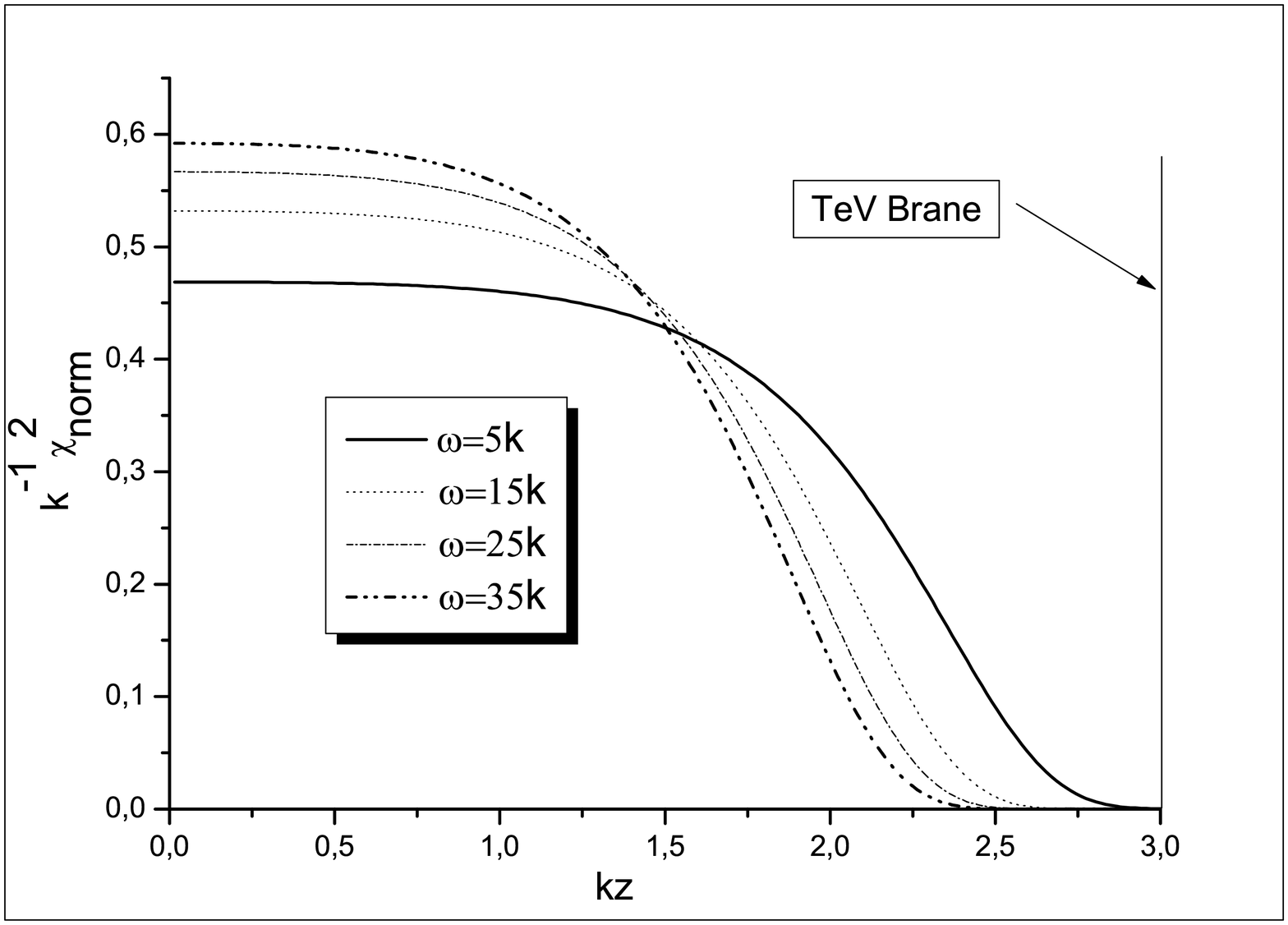}
\end{center}
\caption{\small{The square of the normalized wave function
$k^{-1}\chi_{norm}^2$ of the zero mode (4D photon) as a function of
$kz$ for $k z_{c}=3$ and $\Delta=0.01$ and $\omega=5k,
\:15k,\:25k,\:35k$. We observe that the value of the wave function,
on the Planck brane, increases with the energy and tends to a
constant value.}} \label{7}
\end{figure}

\subsection{Zero mode}
In Fig.\ref{5} and Fig.\ref{6} we have plotted the square of the
normalized wave function measured in units of $k$
($k^{-1}\chi_{norm}^2$) for several values of the energy $\omega$,
assuming that the values of the parameters $\Delta$ and $k z_c$ are
kept fixed. In particular, in Fig.\ref{6}, we observe that for small
energies within the perturbative region of $\omega$
($\omega<\omega_f$) the wavefunction of the photon is almost
constant. This is expected as in the case of RS-model ($\delta h=0$
or $\Delta=0$) the wavefunction can be obtained analytically and it
is constant, see Eq. (\ref{eigen1}). As the energy of the photon
increases the value of the wave function on the TeV brane
($\chi_{norm}(z_c)$) decreases, see Fig.\ref{6}. For
$\omega=\omega_f$ the value of the wave function on the TeV brane is
half of its value at $\omega\simeq 0$, while for larger values of
$\omega$ ($\omega>\omega_f$) it tends rapidly to zero. On the other
hand, we see that the wave function on the Planck brane increases.
Especially, in Fig. \ref{7} we have plotted the wavefunction of the
photon for even larger values of $\omega$, deep in the
nonperturbative sector of the theory. It seems that the wavefunction
on the Planck brane tends to take a limiting value, quite larger
than that for $\omega\simeq 0 $.

If we take into account that the "effective" coupling constant of
the zero mode (4D photon), with matter localized on the TeV brane,
is proportional to $\chi_{norm}(k z_c)$ (see
\cite{Davoudiasl:1999tf, Pomarol:1999ad}), we conclude that photons
with very high energies $\omega>>\omega_f$ tend to decouple from
matter which is localized on the TeV brane. Note, that a similar
behavior has been observed for the massive KK modes of the 5D
photon, which happens for even larger energies, as we see in the
next section. On the other hand, in the case of Planck brane the
coupling of the photon with matter increases with the energy, and
tends to an asymptotic value.

\subsection{First KK excitation}

\begin{figure}[h]
\begin{center}
\includegraphics[scale=0.4,angle=0]{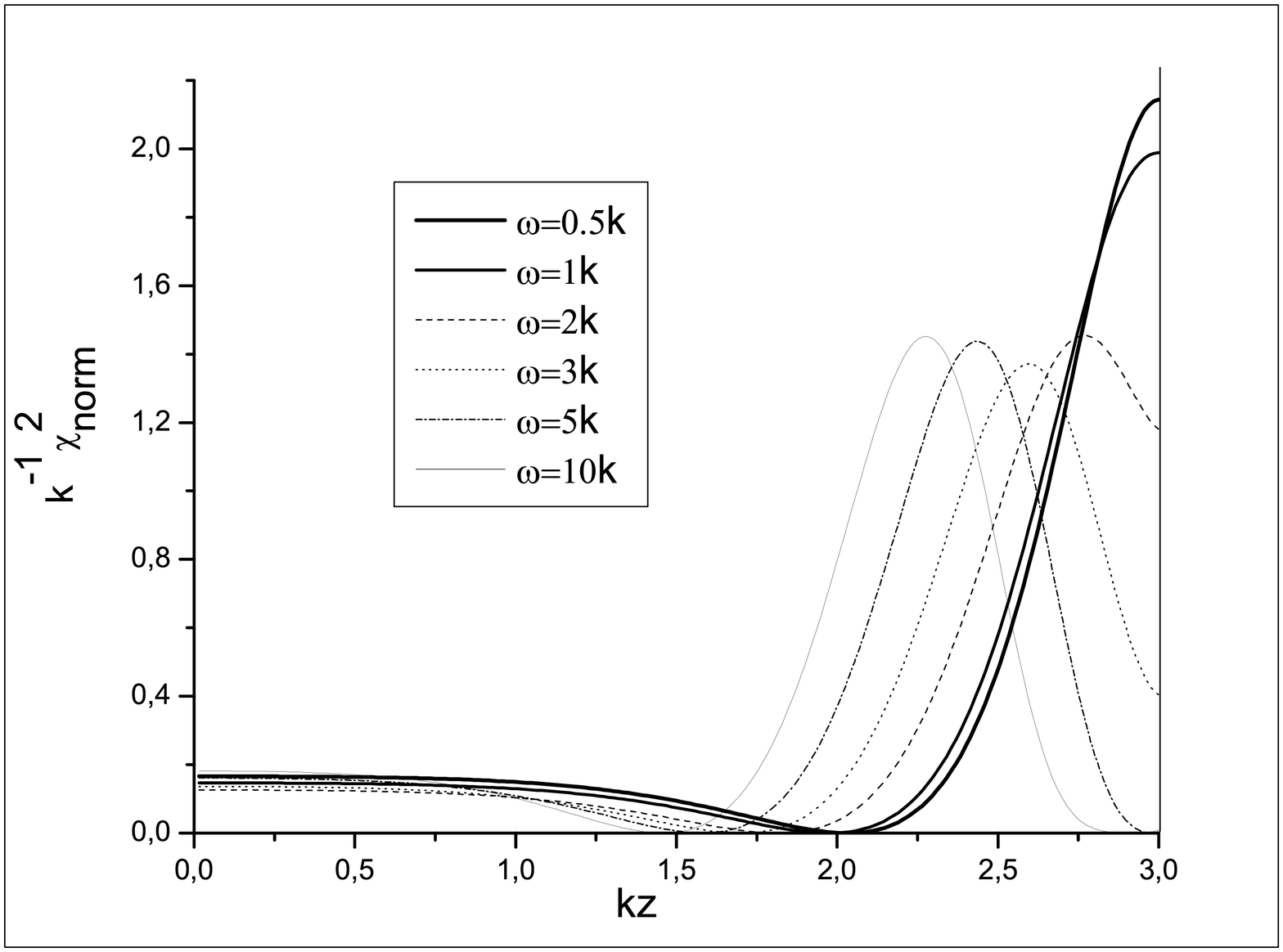}
\end{center}
\caption{\small{The square of the normalized wave function
$k^{-1}\chi_{norm}^2$ of the 1KK mode as a function of $kz$ for $k
z_{c}=3$ and $\Delta=0.01$ and $\omega=0.5k, \:1k,\:2k,\:3k,\:5k,
\:10k$. We see that the value of the wave function on the TeV brane
tends to zero as the energy of the 1KK-mode increases. On the other
hand, on the Planck brane, we observe that the value of the wave
function is almost constant.}} \label{8}
\end{figure}

\begin{figure}[h]
\begin{center}
\includegraphics[scale=0.4,angle=0]{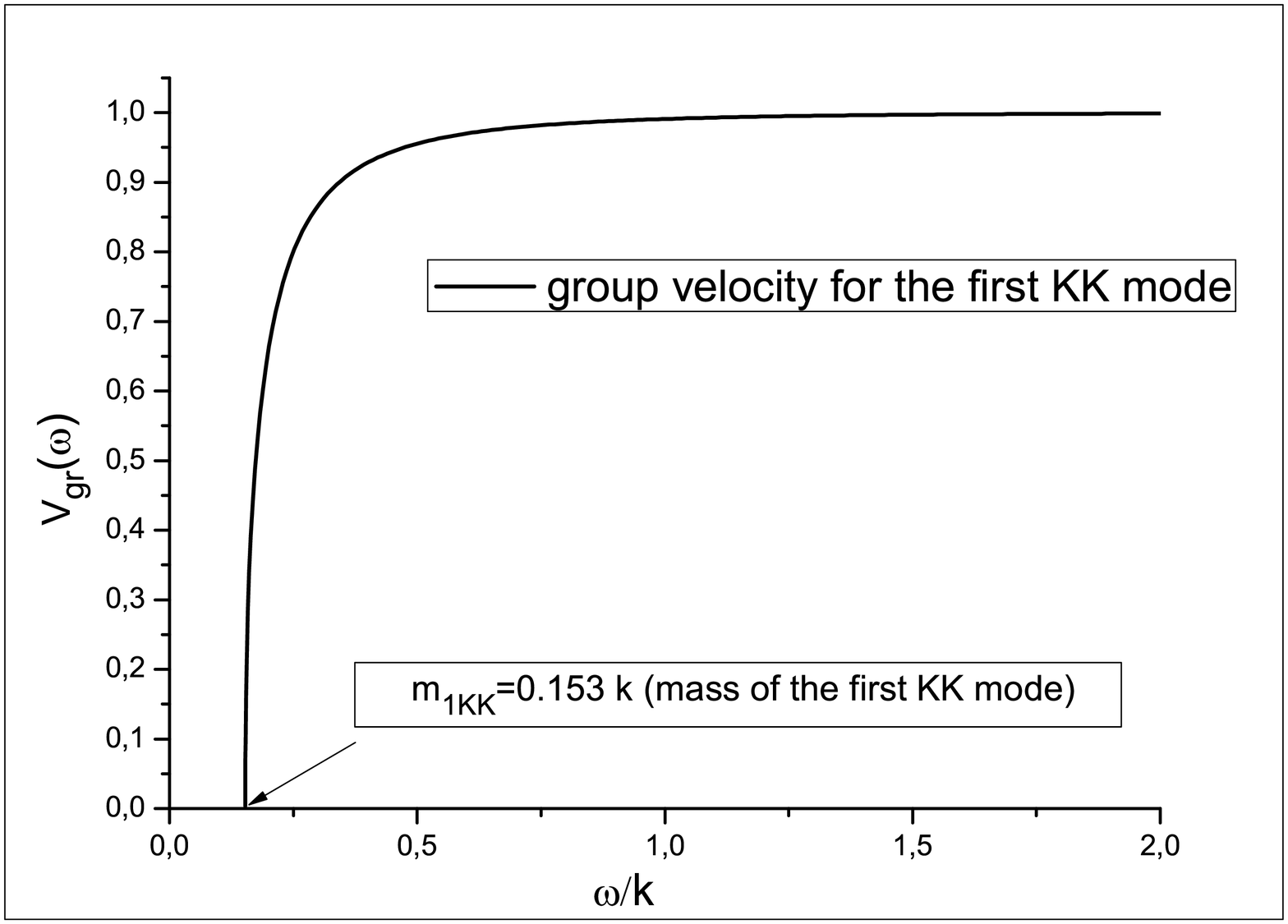}
\end{center}
\caption{\small{The group velocity $V_{gr}(\omega)$ as a function of
energy in the case of the first KK mode, for $kz_c=3$ and
$\Delta=0.001$. Note that the group velocity approaches the unity
remaining always smaller than the group velocity of the zero mode.
We also see in this figure that the energy range has a lower bound
which is identified to the mass of the 1KK mode.}} \label{9}
\end{figure}

In this section we examine the wave function and the group velocity
of the first KK mode. In particular, in Fig. \ref{8}, we see that
the projection of the wave function on the TeV brane decreases as
the energy of of the 1KK mode increases, and for quite large
$\omega$ the value of the normalized wave function becomes almost
zero on the TeV brane. This means that for comparatively large
energies where Lorentz violation effects become significant the 1KK
mode tends to decouple from matter, which is localized on TeV brane.
Note, that a similar behavior was obtained for the zero mode in the
previous section. In the case of higher KK excitations a similar
behavior is expected. On the other hand the projection of the wave
function on the Planck brane is almost constant independently from
the energy $\omega$.

The 1KK mode is a massive particle, as for zero momentum
$\textbf{p}$ the energy $\omega$ takes a nonnegative value
$\omega=m_{1KK}$ ($\neq0$). In Fig. \ref{8}, for $\Delta=0.01$ and
$kz_c=3$, we see that $m_{1KK}=0.153 k$, as it is the lower energy
which is obtained for $\textbf{p}=0$ which corresponds to the
inertial mass of the particle. Note that in the case of $\Delta=0$,
where we can use the formula
\begin{equation}
m^{(0)}_{n}=x_{n}\:k\:e^{-kz_{c}}, \: n=1,2,3,... \label{eigen}
\end{equation}
that gives the masses of the KK excitations, where $x_{n}$ are the
roots of the zeroth order Bessel function $J_{0}(x_n)=0$. For $n=1$
we obtain that $m_{1KK}=0.120 k$. However we can use first order
perturbation theory to correct this value (see ref.
\cite{Farakos:2008rv}), and now we obtain that $m_{1KK}=0.162k$
which is close to the value $m_{1KK}=0.153 k$ that is obtained
nonperturbatively. It is worth noting that in the realistic case
$kz_c=37$ and $\Delta\sim 10^{-8}<<1$ these differences are expected
to be more suppressed, and obviously are not detectable in the
current high energy experiments, for example in LHC.

As the 1KK mode is a massive particle the phase velocity is not
suitable to describe its motion. For this reason in Fig, \ref{9} we
have plotted the group velocity of the particle as a function of
$\omega/k$. We see that the energy has a lower bound which
characterize the inertial mass of the particle as we have also
explained in the previous paragraph. We also see that the group
velocity is always smaller than unity, which is the standard
velocity of light in the tree level of our model, and tends rapidly
to this value $V_{gr}=1$ as the energy $\omega$ increases. Finally,
we would like to note that in our model the group velocity of the
zero mode, even if it becomes smaller than unity as we see in Fig.
\ref{4}, is always larger than the group velocity of the first KK
mode.

\section{Discussion}

We examined a two brane model where the 5D Lorentz invariance is
spontaneously broken due to the nonstandard vacuum of a
five-dimensional charged black hole. In this framework we found a
mechanism which produces an energy dependent vacuum refractive
index, assuming that photons can freely move in the bulk, in
contrast to the conventional brane world hypothesis. As perturbation
theory was examined extensively in a previous work, in this paper we
focused to the nonperturbative case by solving numerically the
eigenvalue problem.

We have mainly studied the phase (and the group) velocity of the
zero mode, 4D photon, and we found that it is in general a
monotonically decreasing function which for very large energies
tends to unity, that is the standard velocity of light at tree level
of our model. Note that a very similar behavior was obtained for the
group velocity of photon, as we can see in Fig. \ref{4} above. On
the other hand, in the case of the first KK mode which is a massive
particle, we found that the group velocity is always smaller than
unity, and it cannot exceed the group velocity of the zero mode in
the high energy limit.

By comparing with perturbation theory we found that there is an
energy $\omega_f$ after which perturbation theory breaks down.
Specifically $\omega_f$ is the inflexion point (see Fig. \ref{5})
where the quadratic dependence on energy terminates and the velocity
tends to a limiting value. One could give an estimate of this point
by comparing with the recent data of the current experiments of
MAGIC \cite{magic,NM}, H.E.S.S \cite{hess} and FERMI \cite{fermi}
telescopes. In the case of our model, which predicts a quadratic
dependence on the energy for the velocity (see Eq. (\ref{vel})), the
stringent bound is set by theMAGIC experiment:
\begin{equation}
V=1-\left(\frac{\omega}{M_{2}}\right)^2, \quad M_{2}\geq 2.6 \times 10^{10} GeV \label{magcon}
\end{equation}
The above restriction was obtained in Ref. \cite{NM} by fitting the
recent experimental data of MAGIC \cite{magic} assuming a quadratic
energy dependence for the photon refractive index. Hence, if we take
the lowest bound for $M_2$ ($M_2=2.6 \times 10^{10} GeV$) we
conclude that the energy $\omega_f$, after which perturbation theory
breaks down, should be quite smaller than $2.6 \times 10^{10} GeV$.
Note that the above upper limit, for the inflexion point $\omega_f$,
if it is expressed in $eV$, gives a value equal to $2.6 \times
10^{19} eV$ which is close (but smaller) to the energy range of the
ultra high energy cosmic rays (particles with astrophysical origin
and energies larger than the GZK limit $7\times 10^{19} eV$).
Accordingly the quadratic dependence of velocity of light from the
energy $\omega$, in the ultra high energy cosmic rays energy region,
is not valid any more. Our analysis, in this region of energies,
shows that the velocity of light is almost independent from the
energy and it has taken a limiting value, which is the velocity of
light at tree level of our model.

In Ref. \cite{sigl} the authors found the following severe
constraint for quadratic dispersion relations:
\begin{equation}
V=1-\frac{1}{2} \xi_{2} \left(\frac{\omega}{M_{PL}}\right)^2, \quad
M_{PL}=10^{19} GeV, \quad \xi_{2}<2.4\times 10^{-7},
\end{equation}
which is due to the lack of observations of photons above the GZK
limit, and appears to be several orders of magnitude stronger than
the bounds of MAGIC observations (compare with Eq. (\ref{magcon})).
However, this constraint presupposes that the quadratic energy
dependence of the velocity of light is valid in the ultra high
energy cosmic ray energies, something which does not happen in our
model as we have mentioned previously. We conclude that the
constraints from the MAGIC observations are the most stringent ones
which can be applied to our model.

\section*{Acknowledgements}

I would like to thank G. Koutsoumbas, N. Mavromatos and P.
Pasipoularides for reading and comment the manuscript. In particular
I wish to thank P. Pasipoularides for useful discussions and help
with mathematica.


\begin{thebibliography}{99}

\bibitem{Mavromatos:2009xg}
  N.~E.~Mavromatos,
  arXiv:0903.0318 [astro-ph.HE].


\bibitem{Farakos:2008rv}
  K.~Farakos, N.~E.~Mavromatos and P.~Pasipoularides,
  JHEP {\bf 0901} (2009) 057
  [arXiv:0807.0870 [hep-th]];
  K.~Farakos, N.~E.~Mavromatos and P.~Pasipoularides,
  arXiv:0902.1243 [hep-th].

\bibitem{antoniadis} I.~Antoniadis,
  Phys.\ Lett.\  B {\bf 246} (1990) 377.

\bibitem{ArkaniHamed:1998rs}
  N.~Arkani-Hamed, S.~Dimopoulos and G.~R.~Dvali,
  Phys.\ Lett.\  B {\bf 429} (1998) 263
  [arXiv:hep-ph/9803315].

\bibitem{Antoniadis:1998ig}
  I.~Antoniadis, N.~Arkani-Hamed, S.~Dimopoulos and G.~R.~Dvali,
  Phys.\ Lett.\  B {\bf 436}, 257 (1998)
  [arXiv:hep-ph/9804398].

\bibitem{Randall:1999ee}
  L.~Randall and R.~Sundrum,
  Phys.\ Rev.\ Lett.\  {\bf 83} (1999) 3370
  [arXiv:hep-ph/9905221].

\bibitem{Randall:1999vf}
  L.~Randall and R.~Sundrum,
  Phys.\ Rev.\ Lett.\  {\bf 83} (1999) 4690
  [arXiv:hep-th/9906064].

\bibitem{Mavromatos:2005yh}
  N.~E.~Mavromatos and E.~Papantonopoulos,
  Phys.\ Rev.\  D {\bf 73} (2006) 026001
  [arXiv:hep-th/0503243].

\bibitem{Mavromatos:2000az}
  N.~E.~Mavromatos and J.~Rizos,
  Phys.\ Rev.\  D {\bf 62} (2000) 124004
  [arXiv:hep-th/0008074];
  N.~E.~Mavromatos and J.~Rizos,
  Int.\ J.\ Mod.\ Phys.\  A {\bf 18} (2003) 57
  [arXiv:hep-th/0205299].

\bibitem{Giovannini:2001ta}
  M.~Giovannini,
  Phys.\ Rev.\  D {\bf 64} (2001) 124004
  [arXiv:hep-th/0107233];
  Class.\ Quant.\ Grav.\  {\bf 23} (2006) L73
  [arXiv:hep-th/0607229];
  Phys.\ Rev.\  D {\bf 75} (2007) 064023
  [arXiv:hep-th/0612104];
  Phys.\ Rev.\  D {\bf 75} (2007) 064023
  [arXiv:hep-th/0612104];
  Phys.\ Rev.\  D {\bf 76} (2007) 124017
  [arXiv:0708.1830 [hep-th]].  


\bibitem{Farakos:2005hz}
  K.~Farakos and P.~Pasipoularides,
  Phys.\ Lett.\  B {\bf 621} (2005) 224
  [arXiv:hep-th/0504014];
  K.~Farakos and P.~Pasipoularides,
  Phys.\ Rev.\  D {\bf 73} (2006) 084012
  [arXiv:hep-th/0602200];
  K.~Farakos and P.~Pasipoularides,
  Phys.\ Rev.\  D {\bf 75} (2007) 024018
  [arXiv:hep-th/0610010];
  K.~Farakos, G.~Koutsoumbas and P.~Pasipoularides,
  Phys.\ Rev.\  D {\bf 76} (2007) 064025
  [arXiv:0705.2364 [hep-th]];
  P.~Pasipoularides and K.~Farakos,
  J.\ Phys.\ Conf.\ Ser.\  {\bf 68} (2007) 012041.

\bibitem{Csaki:2000dm}
  C.~Csaki, J.~Erlich and C.~Grojean,
  Nucl.\ Phys.\  B {\bf 604} (2001) 312
  [arXiv:hep-th/0012143].

\bibitem{Cline:2001yt}
  J.~M.~Cline and H.~Firouzjahi,
  Phys.\ Rev.\  D {\bf 65} (2002) 043501
  [arXiv:hep-th/0107198].

\bibitem{Cline:2003xy}
  J.~M.~Cline and L.~Valcarcel,
  JHEP {\bf 0403}, 032 (2004)
  [arXiv:hep-ph/0312245].



\bibitem{Davoudiasl:1999tf}
  H.~Davoudiasl, J.~L.~Hewett and T.~G.~Rizzo,
  Phys.\ Lett.\  B {\bf 473} (2000) 43
  [arXiv:hep-ph/9911262].


\bibitem{Pomarol:1999ad}
  A.~Pomarol,
  Phys.\ Lett.\  B {\bf 486} (2000) 153
  [arXiv:hep-ph/9911294].

   \bibitem{magic} J.~Albert {\it et al.},
  Astrophys.\ J.\  {\bf 669} (2007) 862
  [arXiv:astro-ph/0702008].

\bibitem{NM}  J.~Albert {\it et al.} [MAGIC Collaboration] and J.~Ellis, N.~E.~Mavromatos,
D.~V.~Nanopoulos, A.~S.~Sakharov and E.~K.~G.~Sarkisyan,
  Phys.\ Lett.\  B {\bf 668} (2008) 253
  [arXiv:0708.2889 [astro-ph]].

  \bibitem{hess} F.~Aharonian {\it et al.},
  Phys.\ Rev.\ Lett.\  {\bf 101} (2008) 170402
  [arXiv:0810.3475 [astro-ph]].

  \bibitem{fermi} A. A. Abdo et al [The Fermi LAT and Fermi GBM collaborations], DOI10.1126/science.1169101
  (Science Express Researce Articles), published online 19 February 2009.

  \bibitem{sigl} M.~Galaverni and G.~Sigl,
  Phys.\ Rev.\  D {\bf 78}, 063003 (2008)
  [arXiv:0807.1210 [astro-ph]].

\end{thebibliography}
\end{document}